\documentstyle[twocolumn,aps,floats,graphicx]{revtex}

\usepackage{color}
\definecolor{Red}{rgb}{1,0,0}
\definecolor{Blue}{rgb}{0,0,1}
\definecolor{Green}{rgb}{0,1,0}
\definecolor{yellow}{cmyk}{0,0,1,0}
\definecolor{Mix}{cmyk}{0.19,0,0.19,0}

\begin{document}

\newcommand{\bec}{\begin{center}}
\newcommand{\ec}{\end{center}}
\newcommand{\be}{\begin{equation}}
\newcommand{\ee}{\end{equation}}
\newcommand{\beqn}{\begin{eqnarray}}
\newcommand{\eeqn}{\end{eqnarray}}
\newcommand{\bet}{\begin{table}}
\newcommand{\ent}{\end{table}}
\newcommand{\bib}{\bibitem}

\newcommand{\red}{\color{Red}}
\newcommand{\blue}{\color{Blue}}
\newcommand{\green}{\color{Green}}
\newcommand{\yellow}{\color{Yellow}}
\newcommand{\mix}{\color{Mix}}


\wideabs{

\title{
Pair condensation and inter-layer coupling in cuprates: pairing on a superlattice 
}

\author{P. S\"ule} 
  \address{Research Institute for Technical Physics and Material Science,\\
Konkoly Thege u. 29-33, Budapest, Hungary,\\
sule@mfa.kfki.hu
}

\date{\today}

\begin{abstract}

\maketitle

According to the recent $c$-axis optical measurements,
we analyze the superconducting state and $c$-axis charge dynamics of cuprates using a charge ordered bilayer superlattice
model in which pairing is supported by inter-layer Coulomb energy gain (potential energy driven 
superconductivity).
The superlattice nature of high-$T_c$ superconductivity is experimentally suggested by the
smallness of the in-plane coherence length $\xi_{ab} \approx 10-30 \AA$ which is comparable with
a width of a $4 \times 4$ to $8 \times 8$ square supercell lattice layer. 
The 2D pair-condensate can be characterized by a charge ordered state
with a "checkboard" like pattern seen by scanning tunneling microscopy.
The $2D \Leftrightarrow 3D$ quantum phase transition of the hole-content at $T_c$,
supported by $c$-axis optical measurements, is also studied.
The pair condensation might lead to the sharp decrease of the normal state $c$-axis anisotropy of the hole content 
and hence to the decrease of inter-layer dielectric screening.
The drop of the $c$-axis dielectric screening can be the primary source of the condensation energy below $T_c$.
We find that a net gain in the electrostatic energy occurs along the $c$-axis, which is proportional to the condensation energy:
$E_c^{3D} \approx E_{cond}$ and is due to inter-layer charge complementarity (charge asymmetry of the boson condensate).
The bilayer model naturally leads to the effective mass of $m^{*} \approx 4 m_e$ found by experiment.
The static $c$-axis dielectric constant $\epsilon_c$ is calculated for various cuprates
and compared with the available experimental data. 
We find correlation between $T_c$ and the inter-layer spacing $d$, $\epsilon_c$
and the coherence area of the condensate.

\end{abstract}
}


\section{Introduction}

 It is more or less generally accepted now that the conventional electron-phonon pairing mechanism 
cannot explain cuprate superconductivity, because as high a transition temperature as
$164 K$ cannot be explained by the energy scale of lattice vibrations without leading to lattice instability \cite{Hirsch}.
It is already well established that much of the physics related to high-temperature superconductivity (HTSC)
is in 2D nature,
one of the basic questions to be answered, however, in the future is whether HTSC is
a strictly 2D phenomenon or should also be described by a 3D theory.
A well known experimental fact is that the zero resistance occurs along the $ab$ plane and the $c$-axis
at the same critical temperature \cite{Plakida} suggesting that there must be inter-layer (IL) coupling involved in the mechanism which drives the system into HTSC.
Another experiment, such as the intercalation on Bi2212 \cite{Choy}, however leads to the
opposite conclusion. Intercalation of I or organic molecules, which expands the unit cell
significantly along the $c$-axis, does not affect $T_c$. 
This finding is against IL coupling and supports low-dimensional theories.
The large anisotropy of the resistivity (and of other transport properties) is again not in favour of 3D theories of HTSC \cite{Plakida}. 
Moreover, Basov {\em et al.} reported $c$-axis optical results and detected a small lowering of kinetic
energy in the IL transport of cuprates \cite{Basov}.
This finding again indicates the importance of IL coupling in high temperature superconductors.
There are a couple of other findings against and supporting the 3D nature of HTSC \cite{Anderson2,PWA,Schneider}.
Most notably the systematic dependence of the transition temperature $T_c$ on the $c$-axis
structure and, in particular, on the number of $CuO_2$ planes in multilayer blocks also
strongly in favour of the 3D character of HTSC. 
It is therefore, a fundamental question whether
at least a weak IL coupling is needed for driving the system to a superconducting (SC) state or a single $CuO_2$
layer is sufficient for superconductivity \cite{Wheatley}.

There has been considerable effort spent on understanding HTSC within the context
of IL coupling mechanism in the last decades in which $c$-axis energy is available as a pairing mechanism
\cite{Mills,Tesanovic,Rajagopal,Xiang}. In other approaches the importance of IL hopping is
emphasized
{\em vs.} the direct IL Coulomb interaction of charged sheets \cite{Anderson2,Wheatley}.
There is another theory providing explanation for HTSC using the general framework of BCS combined
with IL coupling \cite{Ihm}.
The so-called IL tunnelling (ILT) theory \cite{Anderson2,Wheatley}, however, is no longer considered
a viable mechanism for SC in cuprates since ILT could provide no more then $1 \%$ of the
condensation energy in certain cuprates \cite{Hirsch,PWA,Tsvetkov}.

 Recently, optical data have been reported for Bi2212 \cite{Molegraaf}, which supports
the quantitative predictions of Hirsch \cite{HirschPC}, in which the carriers lower their
kinetic energy upon pair formation (kinetic-energy-driven superconductors) and hence
this is the largest contribution to the condensation energy. 
Anyhow, the role of direct IL Coulomb interaction is still not ruled out as a possible
explanation for HTSC, especially if we take into account that no widely accepted
theory is available until now which accounts for the apparent 3D nature of HTSC \cite{Anderson2,PWA}.

  In this paper we propose a simple phenomenological model for explaining the 3D character of HTSC
in cuprates supported by calculations.
We would like to study the magnitude of direct Coulomb interaction
between charge ordered square superlattice layers as a possible source of pairing interaction.
Our intention is to understand HTSC within the context of an IL Coulomb-mediated mechanism.
Of particular relevance to our investigation 
are the doping, multilayer, and pressure dependence of $T_c$ in terms of a charge ordered  {\em superlattice nature} of pair condensation,
which is often neglected in the past.
On the other hand the size of a characteristic superlattice can directly be related to the in-plane coherence length $\xi_{ab}$ of cuprates, which 
is proportional to the linear size of the pair condensate (real space pairing) of the superconducting island in the {ab}-plane \cite{Tinkham}.
The IL charging energy we wish to calculate would then depend on the IL spacing ($d$), the IL dielectric
constant $\epsilon_c$, the hole content $n_h$ and the size of the superlattice.
A large body of experimental data are collected which support our model.
Finally we calculate the static $c$-axis dielectric constant $\epsilon_c$ for various cuprates
which are compared with the available experimental observations.

 It is commonly accepted that charge carriers are mainly confined to the 2D $CuO_2$ layers
and their concentration is strongly influenced by the doping agent via hole
doping (both by chemical and field effect doping \cite{Schon}). Holes (no charge and spin at a lattice site) in the 2D $CuO_2$ layers are the key superconducting elements in high temperature superconductivity (HTSC). A characteristic feature of many high temperature superconductors (HTSCs) is the
optimal hole content value of $n_h \approx 0.16e$ per $CuO_2$ layer at optimal doping in the normal state (NS) \cite{Zhang}.
{\em Our starting hypothesis is that
the hole content goes through a reversible $2D \Leftrightarrow 3D$ quantum phase transition 
at $T_c$ in layered cupper oxides.}
Optical studies on the $c$-axis charge dynamics reveals this phenomenon: the $c$-axis reflectance is nearly insulating
 in the NS but below $T_c$ is dominated by the Josephson-like plasma edge \cite{Basov}.
Below $T_c$ a sharp reflectivity edge is found at very low frequencies (lower then the superconducting gap) for a variety of cuprates
arising from the carriers condensed in the SC state to the $ab$-plane and from the onset
of coherent charge transport along the $c$-axis \cite{Tamasku,Tajima,Katz,Basov2,Motohashi}.
Band structure calculations predict an appreciable $c$-axis dispersion of bands close
to the Fermi surface and thus an anisotropic three-dimensional metallic state \cite{Pickett}.
Other first principles calculations indicate that
above $T_c$ the hole charge $n_h$  is charge transferred to the doping site ($2D \rightarrow 3D$ transition) \cite{Gupta,Sule}.
Furthermore, we expect that
below $T_c$ the hole-content condenses
to the sheets forming {\em anti-hole regions} (hole-content charge at a lattice site, $3D \rightarrow 2D$ transition). 
An anti-hole corresponds to an excess charge condensed to a hole lattice site in the
sheet below $T_c$.
Therefore, in the normal state hole doping is the
dominant charge transfer mechanism while in the superconducting (SC) state the opposite is true: the IL hole-charge is transferred back to the $CuO_2$ planes (charging of the sheets).
The latter mechanism can be seen by the measurement of transport properties as a function
of temperature \cite{Plakida,Ando,Semba,Yamamoto} if we assume that the increase in the density of
in-plane free carriers below $T_c$ is due to pair condensation.
The sharp temperature dependence of the $c$-axis dielectric constant $\epsilon_c$
and optical conductivity \cite{Kitano} 
seen in many cuprates and in other perovskite materials also raises the possibility of
a $2D \Leftrightarrow 3D$ charge density condensation mechanism at $T_c$ \cite{Plakida,diel_t}.
Therefore, the pair condensation can be described by an anisotropic 3D condensation mechanism,
and by a doping induced 2D-3D dimensional crossover \cite{Schneider}.

\begin{figure}

\setlength{\unitlength}{0.07in}
\begin{picture}(20,20)(0,5)
\linethickness{0.35mm}
  \multiput(0,0)(5,0){5}{\line(0,1){20}}
  \multiput(0,0)(0,5){5}{\line(1,0){20}}
  \put(0,0){\circle*{5}}
  \put(0,5){\circle{3}}
  \put(0,10){\circle*{5}}
  \put(0,15){\circle{3}}
  \put(0,20){\circle*{5}}
  \put(5,0){\circle{3}}
  \put(10,0){\circle*{4}}
  \put(15,0){\circle{3}}
  \put(20,0){\circle*{4}}
  \put(5,5){\circle*{4}}
  \put(10,5){\circle{3}}
  \put(15,5){\circle*{4}}
  \put(20,5){\circle{3}}
  \put(5,10){\circle{3}}
  \put(10,10){\circle*{4}}
  \put(15,10){\circle{3}}
  \put(20,10){\circle*{4}}
  \put(5,15){\circle*{4}}
  \put(10,15){\circle{3}}
  \put(15,15){\circle*{4}}
  \put(20,15){\circle{3}}
  \put(5,20){\circle{3}}
  \put(10,20){\circle*{4}}
  \put(15,20){\circle{3}}
  \put(20,20){\circle*{4}}
\end{picture}

\setlength{\unitlength}{0.07in}
\begin{picture}(20,20)(-25,-15)
\linethickness{0.35mm}
  \multiput(0,0)(5,0){5}{\line(0,1){20}}
  \multiput(0,0)(0,5){5}{\line(1,0){20}}
  \put(0,0){\circle{3}}
  \put(0,5){\circle*{4}}
  \put(0,10){\circle{3}}
  \put(0,15){\circle*{4}}
  \put(0,20){\circle{3}}
  \put(5,0){\circle*{4}}
  \put(10,0){\circle{3}}
  \put(15,0){\circle*{4}}
  \put(20,0){\circle{3}}
  \put(5,5){\circle{3}}
  \put(10,5){\circle*{4}}
  \put(15,5){\circle{3}}
  \put(20,5){\circle*{4}}
  \put(5,10){\circle*{4}}
  \put(10,10){\circle{3}}
  \put(15,10){\circle*{4}}
  \put(20,10){\circle{3}}
  \put(5,15){\circle{4}}
  \put(10,15){\circle*{3}}
  \put(15,15){\circle{3}}
  \put(20,15){\circle*{4}}
  \put(5,20){\circle*{4}}
  \put(10,20){\circle{3}}
  \put(15,20){\circle*{4}}
  \put(20,20){\circle{3}}
\end{picture}

\setlength{\unitlength}{0.07in}
\begin{picture}(12,12)(0,-2.6)
\linethickness{0.35mm}
  \multiput(0,0)(5,0){5}{\line(0,1){20}}
  \multiput(0,0)(0,5){5}{\line(1,0){20}}
  \put(0,0){\circle*{4}}
  \put(0,5){\circle*{4}}
  \put(0,10){\circle*{4}}
  \put(0,15){\circle*{4}}
  \put(0,20){\circle*{4}}
  \put(5,0){\circle*{4}}
  \put(10,0){\circle*{4}}
  \put(15,0){\circle*{4}}
  \put(20,0){\circle*{4}}
  \put(5,5){\circle*{4}}
  \put(10,5){\circle*{4}}
  \put(15,5){\circle*{4}}
  \put(20,5){\circle*{4}}
  \put(5,10){\circle*{4}}
  \put(10,10){\circle*{4}}
  \put(15,10){\circle*{4}}
  \put(20,10){\circle*{4}}
  \put(5,15){\circle*{4}}
  \put(10,15){\circle*{4}}
  \put(15,15){\circle*{4}}
  \put(20,15){\circle*{4}}
  \put(5,20){\circle*{4}}
  \put(10,20){\circle*{4}}
  \put(15,20){\circle*{4}}
  \put(20,20){\circle*{4}}
\end{picture}

\setlength{\unitlength}{0.07in}
\begin{picture}(12,12)(-25,-15)
\linethickness{0.35mm}
  \multiput(0,0)(5,0){5}{\line(0,1){20}}
  \multiput(0,0)(0,5){5}{\line(1,0){20}}
  \put(0,0){\circle{3}}
  \put(0,5){\circle{3}}
  \put(0,10){\circle{3}}
  \put(0,15){\circle{3}}
  \put(0,20){\circle{3}}
  \put(5,0){\circle{3}}
  \put(10,0){\circle{3}}
  \put(15,0){\circle{3}}
  \put(20,0){\circle{3}}
  \put(5,5){\circle{3}}
  \put(10,5){\circle{3}}
  \put(15,5){\circle{3}}
  \put(20,5){\circle{3}}
  \put(5,10){\circle{3}}
  \put(10,10){\circle{3}}
  \put(15,10){\circle{3}}
  \put(20,10){\circle{3}}
  \put(5,15){\circle{3}}
  \put(10,15){\circle{3}}
  \put(15,15){\circle{3}}
  \put(20,15){\circle{3}}
  \put(5,20){\circle{3}}
  \put(10,20){\circle{3}}
  \put(15,20){\circle{3}}
  \put(20,20){\circle{3}}
\end{picture}

\vspace{-1.5cm}
\caption{\small Upper panels: The charge odered state of the type of a "checkerboard" of the hole-anti-hole condensate.
Opened and filled circles denote the holes with a charge of $q_i=+0.16e$ and anti-holes ($q_j=-0.16e$),
respectively in the $5 \times 5$ square lattice layer model.
Note that the left panel accomodates $13$ anti-holes which corresponds to
$2e+0.16e$ charge,. The right panel contains $12$ anti-holes corresponding
to $2e-0.16e$ charge.
The two lower panels correspond to the antiferromagnetic insulating state (left)
and to the holed-doped system (normal state, right).
}
\end{figure}
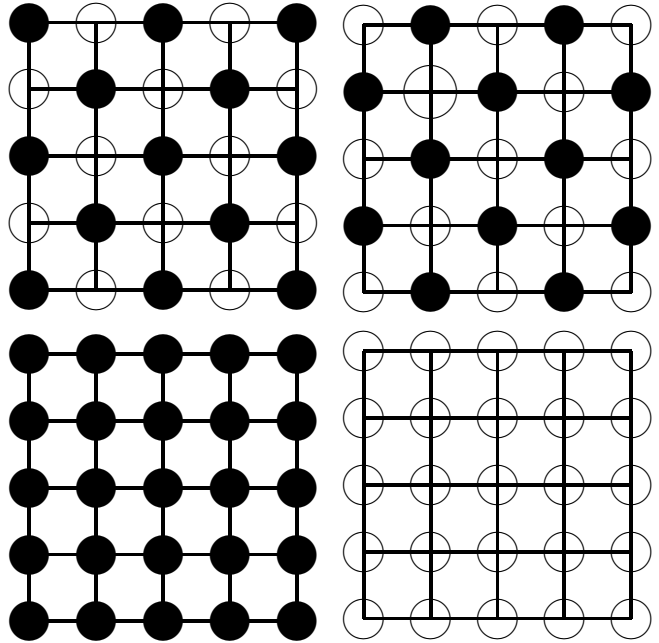

 In any naive model of electron pairing in cuprates the Coulomb repulsion is troublesome.
When the pairs of charged carriers are confined to the 2D sheets, naturally a net self-repulsion
of the pair condensate occurs. Although short-range Coulomb screening in the dielectric crystal
can reduce the magnitude of the repulsion but it is insufficient to cancel
Coulomb repulsion \cite{Friedberg91}. 
Spatial separation should reduce the interaction strength, but even at $14 \AA$
 $e^2/r \approx 1 eV$ if unscreened.
One can assume capacitative effect between the $CuO_2$
planes: The charged boson condensate in one plane is stabilized by a deficiency of that charge
on another plane \cite{Friedberg91}. 
The various forms of the capacitor model are associated with the 3D character of HTSC, considering the inter-layer charge reservoir as a dielectric medium \cite{Friedberg91}.
 The inter-layer charging energy might be insufficient to stabilize the self-repulsion of the holes
in one plane and the self-repulsion of the charge condensate on another plane in the capacitor model.
Also, the superconducting properties of the hole-rich and nearly hole-free sheets would be different which is not verified by experimental techniques.

 Our intention is to combine the 2D and 3D nature of HTSC.
  Therefore, we propose a phenomenological model in which the charge distribution of the planes is polarized 
in such a way that holes and anti-holes (hole-electron pair) are 
phase separated within each of the sheets leading to a charge ordered state.
We also study the magnitude of IL coupling (direct IL Coulomb interaction) between 2D charge ordered superlattices.

   {\em Charging of the sheets}:
 The $CuO_2$ layers carry negative charge obtained from the charge reservoir
even in the insulating stochiometric materials, e.g. in the infinite layer compound
$CaCuO_2$ each layer is charged by $2e$ charge donated by the $Ca$ atoms.
In $(CuO_2)^{2-}$ then the electron configuration of $Cu$ and $O$ is $3d^{10} 4s^1$ and 
$2p_z^2$. 
Unit $(CuO_2)^{2-}$ is typical of any undoped (stochiometric) cuprates and {\em ab initio} calculations
provide approximately the charge state $Cu^{1.5+}O_2^{3.5-}$ \cite{Plakida,Pickett}
which is due to charge redistribution between Cu and O.
The $(CuO_2)^{2-}$ plane itself is antiferromagnetic and insulating, e.g. there are no holes
in the "overcharged" layers.
Upon e.g. Oxygen doping, however, the doping charge $n_h$ is transferred along the $c$-axis
to the doping site, since the doping atom exhibits a rare gas electron 
structure ($O^{2-}$) \cite{Gupta,Sule}.

\begin{figure}

\setlength{\unitlength}{0.05in}
\begin{picture}(30,30)(-18,4)
\linethickness{0.55mm}
  \put(0,0){\line(1,2){5.0}}
  \put(20,0){\line(1,2){5.0}}
  \put(0,20){\line(1,2){5.0}}
  \put(20,20){\line(1,2){5.0}}
  \put(5,0){\line(1,2){5.0}}
  \put(10,0){\line(1,2){5.0}}
  \put(15,0){\line(1,2){5.0}}
  \put(5,20){\line(1,2){5.0}}
  \put(10,20){\line(1,2){5.0}}
  \put(15,20){\line(1,2){5.0}}
  \put(20,0){\line(0,1){20.0}}
  \put(0,0){\line(0,1){20.0}}
  \put(0,20){\line(1,0){20.0}}
  \put(0,0){\line(1,0){20.0}}
  \put(5,30){\line(1,0){20.0}}
  \put(5,10){\line(0,1){20.0}}
  \put(25,10){\line(0,1){20.0}}
  \put(5,10){\line(1,0){20.0}}
  \put(2,3){\line(1,0){20.0}}
  \put(2.5,5.5){\line(1,0){20.0}}
  \put(4,8.){\line(1,0){20.0}}
  \put(1.3,22.5){\line(1,0){20.0}}
  \put(2.3,25.0){\line(1,0){20.0}}
  \put(3.5,27.5){\line(1,0){20.0}}
  \put(0,0){\circle*{3}}
  \put(5,0){\circle{3}}
  \put(10,0){\circle*{3}}
  \put(15,0){\circle{3}}
  \put(20,0){\circle*{3}}
  \put(0,20){\circle{3}}
  \put(5,20){\circle*{3}}
  \put(10,20){\circle{3}}
  \put(15,20){\circle*{3}}
  \put(20,20){\circle{3}}
  \put(1.6,3){\circle{3}}
  \put(6.3,3){\circle*{3}}
  \put(11.5,3){\circle{3}}
  \put(16.5,3){\circle*{3}}
  \put(21.3,3){\circle{3}}
  \put(2.5,5.5){\circle*{3}}
  \put(7.5,5.5){\circle{3}}
  \put(12.5,5.5){\circle*{3}}
  \put(17.5,5.5){\circle{3}}
  \put(22.5,5.5){\circle*{3}}
  \put(4,8){\circle{3}}
  \put(8.6,8){\circle*{3}}
  \put(13.8,8){\circle{3}}
  \put(18.8,8){\circle*{3}}
  \put(23.8,8){\circle{3}}
  \put(4.8,10.5){\circle*{3}}
  \put(9.8,10.5){\circle{3}}
  \put(14.8,10.5){\circle*{3}}
  \put(19.8,10.5){\circle{3}}
  \put(24.8,10.5){\circle*{3}}
  \put(1.3,22.5){\circle*{3}}
  \put(6.3,22.5){\circle{3}}
  \put(10.8,22.5){\circle*{3}}
  \put(16.4,22.5){\circle{3}}
  \put(21.4,22.5){\circle*{3}}
  \put(2.3,25.0){\circle{3}}
  \put(7.3,25.0){\circle*{3}}
  \put(12.3,25.0){\circle{3}}
  \put(17.3,25.0){\circle*{3}}
  \put(22.2,25){\circle{3}}
  \put(3.3,27.5){\circle*{3}}
  \put(8.5,27.5){\circle{3}}
  \put(13.2,27.5){\circle*{3}}
  \put(18.3,27.5){\circle{3}}
  \put(23,27.5){\circle*{3}}
  \put(4.4,30.0){\circle{3}}
  \put(9.4,30.0){\circle*{3}}
  \put(14.4,30.0){\circle{3}}
  \put(19.4,30.0){\circle*{3}}
  \put(24.4,30.0){\circle{3}}
\end{picture}

\vspace{1cm}
\caption{\small The charge ordered state of the hole-anti-hole condensate
in the bilayer $5 \times 5$ superlattice model. Note the charge asymmetry between the adjacent layers which can be reflected by a double "checkerboard" with asymmetrical pattern. The bilayer can accomodate a pair of boson condensate ($\sim 4e$)}

\end{figure}

  Our basic assumption is that in the SC state every second hole with the charge of $+n_h$ is filled up
by the doping charge $n_h$ due to the {\em condensation} of the hole-content $n_h$ to the sheets.
The electrostatic field of the doping site or the external field of the field-effect transistor {\em confines} \cite{Schon} the anisotropic hole-content to the
{\em anti-hole} sites of the sheet leading to a charge ordered
state (FIG. 1) below $T_c$.
Briefly, the $c$-axis anisotropy of the hole content is strongly temperature dependent in cuprates.
{\em The phase separation of the holes and anti-holes (leading to a charge ordered state) is stabilized by the intra- and inter-layer hole-anti-hole interactions.}
Basically the phase separation of holes and anti-holes in the planes is the manifestation of strong Coulomb
correlation in the SC state.
The Coulomb repulsion arising within the anti-hole sites is suppressed by the
hole-anti-hole interactions.

\section{The superlattice model}
  
 We propose to examine the following superlattice model of pair condensation:
 A pair of charge carriers ($2e$) is distributed over $2/0.16=12.5$ $CuO_2$ unit cells in a square
lattice layer if the $2e$ pair is composed of the hole charge $n_h \approx 0.16/CuO_2$.
However, allowing the phase separation of hole-anti-hole pairs, every second unit cell is occupied by
$-0.16e$ (anti-hole), and the rest is empty (holes, $+0.16e$), therefore we have $25$ unit cells for a condensed pair of charge 
carriers
 (Fig 1., that is the unit lattice of the pair condensate in the $5 \times 5$ supercell model).  
Therefore, $2 n_h =-0.32e$ hole charge condenses to every second hole forming anti-hole sites.
The remarkable feature is that the size of the $5 \times 5$ condensate (four lattice spacings, $4 a \approx 15.5 \AA$) is comparable with the measured small coherence
length $\xi_{ab}$ of single-layer cuprates ($\xi_{ab} \sim 10 \AA$ to $20 \AA$) \cite{Plakida,Tinkham,Dagotto,Stinzingen}.
$\xi_{ab}$ can directly be related to the characteristic size of the wave-pocket of the Cooper pair (coherence area) \cite{Plakida,Tinkham}.
The supercell model can be applied not only for $n_h=0.16e$ but also for the entire doping regime.
Our expectation is that the charge ordered state of the $5 \times 5$ model given in Fig. 1 can be an effective 
model state for describing the SC state.
An important feature of this model is the {\em charge separation $dq$} in the charge ordered state,
where $+0.16e$ and $-0.16e$ partial boson charges are localized alternatively ($dq=0.32e$).
The hopping of a partial charge from the anti-hole sites to the holes can reduce the magnitude of the charge separation $dq$ leading to the extreme case when
$dq=0$ which is nothing else then the $(CuO_2)^{2-}$ antiferromagnetic insulating state. 
Therefore characteristic quantity of the SC state $dq$ is directly related to the hole-content seen in cuprates in the
NS as a function of doping. 

 The hopping of anti-holes in such a charge ordered lattice layer might lead to SC without
considering any lattice vibration effects when $dq$ is in the optimal regime.
Due to the charge redistribution mentioned above net repulsion occurs between the portions of the $2e$ charge within different sites in the SC state
(intra $CuO_2$ repulsion), and attraction occurs between the adjacent $CuO_2$ units with
opposite charges (inter $0.16-...0.16+$ interaction).
Also, charging energy gain occurs out-of-plane due to inter-layer attraction ($E_{el}^c$).
Below $T_c$ the charged-ordered state becomes stable compared with the
competing phase of the NS supported by IL charging energy.
Holes and anti-holes are placed in such a way in adjacent layers to maximize $E_{el}^c$.
Therefore, holes in one of the layers are always covered by anti-holes in the other layer and {\em vice versa}
(FIG. 2, inter-layer {\em electrostatic complementarity}, bilayer $5 \times 5$ model).
An important feature is then that
the boson condensate can be described by an inter-layer {\em charge asymmetry}.
The IL coupling of boson-boson pairs in the bilayer $5 \times 5$ model naturally suggests the
effective mass of charge carriers $m^{*} \approx 4 m_e$, as it was found by measurements \cite{Stasio,krusin}.
Notable feature of the antisymmetric $5 \times 5$ bilayer model is that the basel plane
contains $13$ anti-holes and $12$ holes while the reverse is true for the adjacent layer: it contains $12$ anti-holes and $13$ holes (FIG. 2).
In the next section we generalize the $5 \times 5$ model to represent a $N \times N$ coherence
area.
The charge ordered state presented in this article can also be studied as a charge density
wave (CDW) condensed on a superlattice.
Using the CDW terminology one has approximatelly two CDWs within a characteristic bilayer
with an amplitude of $\sim 0.16e$.
This system might not be insulating if we assume high rate of intersite hopping
and low rate of IL hopping of the charge carriers.
Then the charge ordered state is stabilized due to the attractive IL 
Coulomb interactions of asymmetrically condensed neighbouring CDWs.

\begin{figure}

\setlength{\unitlength}{0.06in}
\begin{picture}(20,20)(-17,2)
\linethickness{0.35mm}
  \multiput(0,0)(5,0){5}{\line(0,1){20}}
  \multiput(0,0)(0,5){5}{\line(1,0){20}}
  \put(0,0){\circle*{2}}
  \put(0,5){\circle{4}}
  \put(0,10){\circle*{2}}
  \put(0,15){\circle{4}}
  \put(0,20){\circle*{2}}
  \put(5,0){\circle{4}}
  \put(10,0){\circle*{2}}
  \put(15,0){\circle{4}}
  \put(20,0){\circle*{2}}
  \put(5,5){\circle*{2}}
  \put(10,5){\circle{4}}
  \put(15,5){\circle*{2}}
  \put(20,5){\circle{4}}
  \put(5,10){\circle{4}}
  \put(10,10){\circle*{2}}
  \put(15,10){\circle{4}}
  \put(20,10){\circle*{2}}
  \put(5,15){\circle*{2}}
  \put(10,15){\circle{4}}
  \put(15,15){\circle*{2}}
  \put(20,15){\circle{4}}
  \put(5,20){\circle{4}}
  \put(10,20){\circle*{2}}
  \put(15,20){\circle{4}}
  \put(20,20){\circle*{2}}
\end{picture}

\vspace{1cm}
\caption{\small The charge ordered state of the normal state.
Opened and filled circles denote the holes with a charge of $q_i \approx +0.32e$ and anti-holes ($q_j \approx -0.16e$),
respectively in the $5 \times 5$ square lattice layer model.
}
\label{fig_ns}
\end{figure}
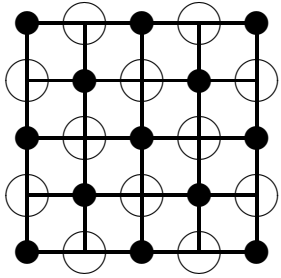

 The following charging model systems can be considered:

 (1) {\em purely hole-doped model}: In each sheet each $CuO_2$ units possese $n_h=+0.16e$ charge.
 The hole charge $n_h$ is charge transfered to the charge reservoir (doping site).
 In this model one can naively expect that the self-repulsion of a hole rich sheet can be quite large. In order to reduce the hole-hole self-repulsion of the sheets, a hopping of $\sim n_h$ between the adjacent hole sites would result in the reduction of the repulsion. As is well known, the hopping of certain amount of charge ($\sim n_h$) between the holes leads to metallic conductance which is typical of hole-doped cuprates (FIG ~\ref{fig_ns}).
Strong Coulomb forces also arise between the negatively charged doping site
and the positive sheets. This interaction may well stabilize the system.
This is the {\em normal state} (NS) of the cuprates. 
Due to the strong $c$-axis anisotropy of the hole charge in the NS 
IL dielectric screening is large in this state because of the increase in the dynamic component
of the IL dielectric permittivity ($\epsilon_c$).

\vspace{3cm}

\vspace{-3cm}

 (2) {\em charged sheet model}: Each of the superconducting sheets ($CuO_2^{2-}$) 
and the charge reservoir in its doping site are charged. 
According to {\em ab initio} calculations \cite{Sule}, the doping oxygen in Hg1201 can get
the excess negative charge either from the charge reservoir or from the planes.
The doping charge or the external field
polarizes the diffuse charge distribution of the sheets. In this way holes and anti-holes
occur in the planes which leads to inter-layer charging energy. 
In the sheets {\em charge patterns} can be formed, e.g. hole-rich and anti-hole rich regions.
Within this model $\pm 0.16e$ charges/$CuO_2$ are placed alternatively in the lattice ($dq=0.32e$).
The pair-binding energy is provided by the self-interaction of the
charge-ordered condensate at low temperature. However, our proposal is to understand
HTSC in such a way that the pair-condensate is stabilized at higher
temperature by the IL-charging energy provided by the antisymmetry of the
pair-condensate charge density between the adjacent layers.

 The $c$-axis dielectric constant of the SC state is reduced to the average value of
the IL dielectric medium due to the pair condensation of the hole charge to the sheets.
No IL screening of the hole charge occurs in this state which may well lead to energy gain
and hence Coulomb induced pairing in the sheets.
Our proposal is that when the system goes to SC state, the conduction band loses its
3D anisotropy due to the 2D pair-condensation of the hole-content which leads to resistivity-free
hole-conductivity in the planes below $T_c$.

\section{The total energy of the pair condensate}

 Those who believe that high-$T_c$ Cooper pairing can be attributed to Coulomb forces
would agree that
the key ingredient in this expression is the energy- and momentum-dependent dielectric function
$\epsilon_c$ of the
solid.
Our intention is to develop a simple working hypothesis in which the IL charge reservoir
provides an average dielectric background and the hole content further enhances
IL dielectric screening when $2D \rightarrow 3D$ transition occurs (hole-doping, the NS).
The dielectric plasma provided by the hole charge results in the dynamic screening
effect of Coulomb interaction which is typical of hole doped cuprates in the NS.
In the opposite case ($3D \rightarrow 2D$, pair condensation) the IL dielectric screening
is nearly reduced to the average background value of the IL ion core spacer. This is what leads to IL energy gain.
The underlying source of the condensation energy is then the energy gain due to the lack of
dynamic screening in the SC state.
The possibility of direct IL hopping of Cooper pairs is not considered within this model as it was ruled out as
important aspect of HTSC \cite{PWA}.

We start from a very general description of our model system using e.g. a
Hamiltonian similar to that is given elsewhere \cite{Tesanovic} or which
can also partly be seen in general text books \cite{Kittel}.
We would like to describe then the
$2D \Leftrightarrow 3D$ condensation of the hole-content 
 using the Hamiltonian
\be
H=\sum_i H_i^{2D}+\sum_{i,j} H_{ij}^{3D},
\label{hamilt}
\ee
where $H_i^{2D}$ is the BCS-type Hamiltonian of the intra-layer condensate.
\beqn
H_i^{2D}=\sum_{k,\sigma} \epsilon_k c_{k \sigma,i}^{\dagger} c_{k \sigma,i}
\\ \nonumber 
+V \sum_{k,k'} c_{k\uparrow,i}^{\dagger} c_{-k\downarrow,i}^{\dagger} c_{k\downarrow,i}
c_{-k\uparrow,i}.
\label{intra}
\eeqn
In Eq. (2) $c_{k\uparrow,i}^{\dagger}$ is the creation operator for electrons in the
$i$th layer, with linear momentum $\bf k$ within the layer and spin $\sigma$,
using the effective-mass approximation $\epsilon_k=\hbar^2 k ^2/2 m^*$. 
$V= -\vert V \vert$ is the attractive inter-layer interaction (3D coupling) and is assumed
to originate from some of the proposed mechanisms \cite{Anderson2,Mills,Tesanovic,Rajagopal}.
The attractive IL term $H_{ij}^{3D}$ given by
\beqn
H_{ij}^{3D}=-t \sum_{k \sigma,\alpha,\beta} c_{k\sigma,i}^{\dagger} c_{k\sigma,j}
+ Y\sum_{k,k'} c_{k \alpha,i}^{\dagger} c_{-k \beta,j}^{\dagger} 
c_{-k \beta,j} c_{k \alpha,i} \\ \nonumber
+W\sum_{k,k',\alpha,\beta} c_{k \alpha,i}^{\dagger} c_{-k \beta,i}^{\dagger}
c_{-k \beta,j} c_{k \alpha,j}.
\eeqn
The first term in Eq. (3) is the direct IL hopping, while the second and third
terms describe IL coupling in a general form.
$t$ is very small in oxide superconductors \cite{PWA,Tesanovic}. $Y$ denotes
the IL coupling assisted by 
direct Coulomb interaction between charged layers (this is of particular
interest in our model).
$W$ is the coupling constant and can arise through the Coulomb interaction
causing inter-band transitions at the Fermi surface to some of the occupied or empty bands
away from the Fermi level, with finite dispersion, however, along the
$c$-axis.
In this paper we consider only the second term in Eq. (3) as the source of
the attractive interaction for pairing and neglect the rest of the inter-layer Hamiltonian
$H_{ij}^{3D}$ ($t \approx 0, W \approx 0$).

  We calculate then the energy of the nearly 2D electron pair condensate in the $CuO_2$ plane for the
bilayer supercell problem. 
It is assumed that the pair condensate behaves as a nearly free electron
gas with $ab$-kinetic energy $E_{kin}^{ab}$ and potential energy, which is mainly its in-plane self-electrostatic
energy ($E_{el}^{ab}$) and the out-of-plane inter-layer interaction energy $E_{el}^c$.
For simplicity, the rest of the electron system is neglected completely.
The external potential of the condensate is also excluded in this model, that is the
lattice-condensate interaction, which is assumed to be negligible in the charged
$(CuO_2)^{2-}$ system (at least its contribution is negligible to the condensation energy).
In other words the ionic background of the planes is screened by the
core and valence electrons of the $(CuO_2)^{2-}$ plane. 
The kinetic energy of the charge condensate is due to the hopping of the charge carriers
between the $CuO_2$ sites within the sheets.
We do not take into account the complications due to ionic heterogeneity, nonpointlike polarization, etc.
The effective Coulomb interaction between spinless point-charges 
may be approximated by the expression $V_{eff}({\bf {r}})=e^2/(4 \pi \epsilon_0 \epsilon_c {\bf {r}})$, where $\epsilon_c$ takes into account phenomenologically the dielectric screening
effect of the IL dielectric medium and confined hole charge.
This kind of a rough approximation has widely used by several groups in the last decade
\cite{Mills,Friedberg91,Stasio,Ariosa}.
The important feature is that the $\underline{c}$-axis dielectric screening ($\epsilon_c$) is nearly reduced
to the static value of the average background dielectric constant in the SC state.
In accordance with this
a sharp temperature dependence of $\epsilon_c(\omega)$ is found in BSCCO by $c$-axis
optical measurements at $T_c$ \cite{Kitano}.
The static out-of-plane dielectric function can be obtained from the sum rule
\be
\epsilon_c=\epsilon_1(0)+\frac{2}{\pi} \int_0^{\infty} \frac{\epsilon_2(\omega')}{\omega'}
{d\omega'},
\ee
where $\epsilon_2(\omega)$ is the dynamic component of $\epsilon_c(\omega)$ \cite{Pines}.

 Although, the $r$-dependence of the screened in-plane electrostatic interaction
between the condensed charges $q_i$ and $q_j$ is not perfectly $1/r$, we
approximate it with the expression $V_{eff}^{ab} \approx e^2/4 \pi \epsilon_0 
\epsilon_{ab} r$ as well, where screening is taken into account implicitly
via $\epsilon_{ab}$.
The lowest eigenvalue of $H$ given in Eq. ~(\ref{hamilt}) is $E=\langle \Psi \vert H \vert \Psi \rangle$, where $\Psi(r_1,r_2)=\sum_{\vec{k}}^{'} g(k) e^{i\vec{k}\cdot\vec{r}_1} e^{i\vec{k}\cdot\vec{r}_2}$
being
the wavefunction of the interacting pair and $g(k)$ is a pair-correlation function
\cite{Tinkham}.
Then the energy of the lowest eigenstate of the condensate in the SC state is approximated by
using a pure dielectric form for the potential energy
\beqn
E=E_{kin}^{ab}+E_{el}^{ab}+E_{el}^c=\frac{\hbar^2}{2 m^{*}}\Delta \Psi(r_1,r_2) ~~~~~~~~~~~~~~~~~&& \\ \nonumber~~~~~~ +\frac{e^2}{4 \pi\epsilon_0} \biggm(\frac{1}{\epsilon_{ab}} \sum_{ij}
\frac{q_i^{(1)} q_j^{(1)}}{r_{ij}^{(1)}} 
~~~~~~~~&& \\ \nonumber
  +\frac{1}{\epsilon_c} \sum_{n=1}^{2} \sum_{m=2}^{\infty} \sum_{ij}^{2 N^2} \frac{q_i^{(n)} q_j^{(m)}}{r_{ij}^{(n,m)}}\biggm),~~~~
\eeqn
where $\hbar$ is the Planck constant, $m^{*} \simeq 4 m_e$ is the effective mass for the holes induced in the half-filled bands \cite{Stasio},
$n,m$ are layer indices,
$r_{ij}^{(1)}$ and $r_{ij}^{(n,m)}$ are the intra-layer and inter-layer point charge distances, respectively.
The most important components are the interactions with $r_{ij}^{(1,2)}$ (bilayer components),
however one has to sum up for the IL interactions with terms $r_{ij}^{(n,m)}$,
where $m=[2,\infty]$. Note that only the interactions of various layers
with the basel bilayer (FIG. 2) are considered up to infinity along the $c$-axis in both directions (up and down).
Interactions within the basel plane is considered only within the coherence area ($N \times N$ superlattice).
Reason for this is very simple: although formally in-plane interactions are incorporated
into Eq. (5), however, later we will see that interactions only along the $c$-axis are
important within the bilayer model of pair-condensation.
Intra-plane interactions of the reference 2D-condensate over the coherence area is beyond the scope
of the present study and should be considered in a near future work.
The in-plane electrostatic screening $\epsilon_{ab}$ is completely separated from
the out-of-plane dielectric screening ($\epsilon_c$).
$q_i, q_j$ are the partial point charges/atoms in the $N \times N$ superlattice model.
\be
q_{i,j}=\pm \frac{2 N_h}{3 N^2}, 
\label{partialcharges}
\ee
where factor $2$ is due to the fact that
every second $CuO_2$ site is occupied by anti-holes, and each of them consists of $3$ atoms
($q_{i,j} \approx \pm 0.053e$ at maximal charge separation).
For the sake of simplicity it is assumed that the charges are equally distributed
among Cu and O atoms within a $CuO_2$ site.
The number of the lattice sites in the characteristic superlattice
is $N^2$ and 
\be
\xi_{ab} \approx (N-1)a_0
\label{xiab}
\ee
where $a_0 \approx 3.88$ \AA is the $ab$ lattice constant.
$N_h=2e$ is the charge of the electron pair.
The anti-hole charges $q_i^{ahole}$ must satisfy the {\em charge sume rule} within a characteristic bilayer over a coherence area $\sim \xi_{ab}^2$
\be
\sum_{i=1}^{N^2} q_i^{ahole} = 4e,
\ee
where $q_i^{ahole}$ represents the anti-hole point charges.
Naturally the charge neutrality $\sum_{ij}^{2 N^2} (q_i^{hole}+q_j^{ahole})=0$ is also required. 
In this paper we study the lattice size $N \times N=5$, which we found
nearly optimal for a variety of cuprates. However, it can be interesting
to study the variation of the "characteristic" lattice size in different
cuprates as a function of various parameters (doping, pressure etc.). 
The kinetic energy of the boson condensate arises from
the hopping of anti-holes (hole-anti-hole exchange) between the adjacent sites within the sheets,
against the electrostatic background of the rest of the hole-anti-hole system. Single hole-anti-hole
exchange is forbidden since extraordinary repulsion occurs which leads to the redistribution
of the entire hole-anti-hole charge pattern. The collective intersite anti-hole hopping results in the
kinetic energy of the condensate.
 If we chose appropriately the in-plane dielectric constant ($\epsilon_{ab} \approx 3.0$ \cite{Molegraaf,epsilonc}), we get the following
result using Eq. (2):
\be
 E_{kin}^{ab}+E_{el}^{ab} \approx 0.
\label{ab}
\ee
Consequently, no net in-plane energy gain is available for HTSC in this model.
In other words we expect no role of pairing induced kinetic energy gain contrary to other
theories \cite{Hirsch,Anderson2,HirschPC} since the kinetic energy of the condensate is
cancelled by the in-plane electrostatic energy.
This is basically the manifestation of the virial theorem ($2T+V=0$) for the $ab$-plane and is
reasonable to expect its validity when weak IL coupling is assumed, where
$E_{kin}^{ab}=2T$ and $E_{el}^{ab}=V$.
If the inter-sheet coupling is not that weak, then kinetic energy gain can occur
in the planes (the violation of the in-plane virial theorem).
It can therefore be the subject of further studies that
both $ab$-plane kinetic energy and the screened $c$-axis IL Coulomb interaction can be
the source of the condensation energy in the SC state.
It is interesting that
Eq.~(\ref{ab}) holds when the measured mass enhancement of $m^{*} \approx 4 m_e$ is used \cite{Stasio}.
We can account for this mass enhancement if 
a pair of boson condensate charge $4e$ is distributed within the bilayer $N \times N$ model.
The basic question is whether the pair condensation ($3D \rightarrow 2D$ transition of the hole charge) 
changes the kinetic energy or not in cuprates, however, is not well understood yet.
For the sake of simplicity we neglect here the role of kinetic energy lowering and
employ Eq.~(\ref{ab}) as it stands.

Taking the energy difference $\Delta E_{tot}= \vert E_{tot}^{NS}-E_{tot}^{SC} \vert$,
the energy gain in the SC state (condensation energy) can be given as follows,
\be
  E_{cond} \approx \Delta E_{el}^c = \vert E_{el}^{c,NS}-E_{el}^{c,SC} \vert.
\label{gain}
\ee
The NS contribution to Eq.~(\ref{gain}) is
$E_{el}^{c,NS} \approx 0$, if IL coupling is screened effectively (large density of the hole
content in the IL space $\leadsto$ large $\epsilon_c(NS)$)
then Eq.~(\ref{gain}) is reduced to 
\be
2 N^2 E_{cond}^{exp} \approx  E_{cond} \approx E_{el}^{c,SC},
\label{gain_sc}
\ee
where the experimental condensation energy $E_{cond}^{exp}$ is usually
given per unit cell, therefore the factor of $2 N^2$ is employed
in order to compare with the calculated $E_{cond}$ for the superlattice
of the coherence area.
Further studies will be needed, however, to understand the effect of $E_{el}^{c,NS}$ on Eq.~(\ref{gain}). 

 The  energy gain in the SC state is provided by the change in the inter-layer charging energy which
is essentially the change of the out-of-plane hole-anti-hole interaction energy below $T_c$.
{\em However, the hole-conductivity in the SC state is strictly 2D phenomenon, no
direct IL hopping of quasiparticles is considered within this model.}
$T_c$ is mainly determined by the inter-plane distance
and by the static $c$-axis dielectric constant (the c-axis component of the dielectric tensor).
{\em An interesting feature of the $N \times N$ bilayer model is then that it is capable of retaining
the 2D character of SC while $T_c$ is enhanced by 3D Coulomb interactions.}


\section{The dielectric constant $\epsilon_c$ within the BCS limit}


The model outlined in the previous section allows us to estimate the static
dielectric constant $\epsilon_c$ for various cuprates.
First we give the brief derivation of the formula starting from the
BCS equation for the gap \cite{Tinkham},
\be
k_{BCS} k_B T_c = \Delta(0),
\ee
where $k_B$ is the Boltzman constant and the gap ratio $k_{BCS} \approx 3.53$ for normal superconductors at weak-coupling limit \cite{Tinkham}.
The BCS estimate of
the condensation energy (the energy gain in the SC state) is proportional, however
to $\sim \Delta_g^2$, therefore we can use the following expression \cite{Tinkham},
\be
E_{cond}=-\frac{1}{2} N(0) \Delta(0)^2 = -\frac{1}{2} N(0) k_{BCS}^2 k_B^2 T_c^2,
\label{cond}
\ee
where $N(0)$ is the density of states at the Fermi surface \cite{Tinkham}.
We recall now Eq. ~(\ref{gain_sc}),
\be
E_{cond} \approx E_{el}^c=\frac{e^2}{4 \pi \epsilon_0 \epsilon_c} \sum_{n=1}^{2} \sum_{m=2}^{\infty} \sum_{ij} \frac{q_i^{(n)} q_j^{(m)}}{r_{ij}^{(n,m)}}
,
\ee
from which $\epsilon_c$ can be derived,
\be
\epsilon_c \approx \frac{e^2}{4 \pi \epsilon_0 E_{cond}} \sum_{n=1}^{2} \sum_{m=2}^{\infty} \sum_{ij} \frac{q_i^{(n)} q_j^{(m)}}{r_{ij}^{(n,m)}}
.
\label{epsc}
\ee
Using Eq. ~(\ref{cond}) $E_{cond}$ can be substituted into Eq. ~(\ref{epsc}) leading to
a new expression,
\be
  \epsilon_c = \frac{2 e^2}{4 \pi \epsilon_0 N(0) k_{BCS}^2 k_B^2 T_c^2} 
\sum_{n=1}^{2} \sum_{m=2}^{\infty} \sum_{ij}^{2 N^2} \frac{q_i^{(n)} q_j^{(m)}}{r_{ij}^{(n,m)}}.
\label{epsc2}
\ee
$N(0)$ can be obtained from first-principles calculations or from specific-heat
measurements. $N(0)$ is connected with the Sommerfeld constant $\gamma$
in the low-temperature electronic specific heat by the relation
\be
N(0)=\gamma [2 \frac{\pi^2}{3} k_B^2 (1+\lambda)]^{-1},
\ee
$\lambda$ is a coupling constant due to strong electron-phonon renormalization.
$\gamma$ can be obtained from specific heat measurements \cite{Plakida} 
\be
1.43 \gamma= \Delta C/T_c,
\ee
where $\gamma$ is usually given in [mJ/mol.$K^2$] therefore $ \sim 2 N^2 \gamma$ must be
used for the coherence area of $\sim \xi_{ab}^2$ (twice for the bilayer).
Finally we get the expression for the static $c$-axis dielectric constant
\be
  \epsilon_c = \frac{\pi e^2 (1+\lambda)}{6 \epsilon_0 k_{BCS}^2 N^2 \gamma T_c^2}
  \sum_{n=1}^{2} \sum_{m=2}^{\infty} \sum_{ij} \frac{q_i^{(n)} q_j^{(m)}}{r_{ij}^{(n,m)}}.
  \label{epsc3}
\ee
The only uknown parameter in Eq. ~(\ref{epsc3}) is $\lambda$ which is in the
range of $\lambda=1-2$ when strong electron-phonon coupling is assumed \cite{Plakida}.

\section{Beyond the BCS limit}

 There are number of evidences are available which suggest that HTSC occurs beyond
the BCS limit. In those materials which contain nearly isolated
single layers, such as $Bi_2Ba_2CuO_{6+\delta}$  \cite{Konstantin} or 
superlattice structures (periodic artificially layered materials) such as O-doped
$(BaCuO_2)_2/(CaCuO_2)_n$ thin films, when $n=1$  \cite{Balestrino} and in
$YBa_2Cu_3O_{7-\delta}/PrBa_2Cu_3O_{7-\delta}$ \cite{Plakida,Li} the critical temperature is 
limited to $T_c \le 30 K$ (below the BCS limit).
Therefore it is worth to explain the enhancement of $T_c$ beyond the BCS limit
assuming other mechanism than the electron-phonon coupling.
Inter-layer Coulomb coupling can be a natural source of the condensation
energy and of HTSC.
Assuming that thermical equilibrium occurs at $T_c$ for the competing
phases of the NS and the SC state the following equation
for the condensation energy can be formulated:
\be
2 N^2  E_{cond}^{exp} \approx k_B T_c \approx E_{el}^{c,SC}
\label{kbtc}
\ee
This expression leads to the very simple formula for the critical temperature
\be
T_c (N,dq,d,\epsilon_c) \approx \frac{e^2}{4 \pi \epsilon_0 \epsilon_c k_B} \sum_{n=1}^{2}\sum_{m=2}^{N_l} \sum_{ij}^{2 N^2} \frac{q_i^{(n)} q_j^{(m)}}{r_{ij}^{(n,m)}}
\label{kbtc2}
\ee
where $N_l$ is the number of layers along the $c$-axis. When $N_l \rightarrow \infty$, bulk $T_c$ is calculated.
$T_c$ can also be calculated for thin films when $N_l$ is finite.
and $\epsilon_c$ can also be derived
\be
\epsilon_c \approx \frac{e^2}{4 \pi \epsilon_0 k_B T_c} \sum_{n=1}^{2}\sum_{m=2}^{N_l} \sum_{ij}^{2 N^2} \frac{q_i^{(n)} q_j^{(m)}}{r_{ij}^{(n,m)}}
,
\label{epsc4}
\ee
where a $c$-axis average of $\epsilon_c$ is computed when $N_l \rightarrow \infty$.
Within this picture
the IL interaction energy of the charge ordered condensate in the SC state
is available as condensation energy.
It is not necessary, however, to assume that $E_{el}^{c,SC}$ is available as pair-binding
energy,
simply the superconducting phase is
stabilized {\em vs.} the NS due to $E_{el}^{c,SC}$.
Therefore $E_{cond}^{HTSC} \approx k_B T_c$ holds because of thermodynamic
reasons and not because of the microscopic mechanism of pairing.

In order to test the validity of our approach we estimate the {\em condensation
energy} and compare with the measured value. According to Tsvetkov {\em et al.} $E_{cond}^{exp}=3.7.10^{-6}$ Hartree/unit cell ($\sim 100 \mu eV/Cu$) for $Tl_2Ba_2CuO_6 (T_c=85 K)$ \cite{Tsvetkov}. For the
bilayer $5 \times 5$ model we have $50 E_{cond}^{exp} \approx 0.19 mH$
($2 N^2=50$) which is really comparable with $k_B T_c=0.27 mH$.
The $6 \times 6$ lattice would give relatively small $\epsilon_c=2.6$, however
$72 E_{cond}^{exp} \approx 0.27 mH$ is in agreement with $k_B T_c$.
 In the case of YBCO ($YBa_2Cu_3O_7$) the $N=4$ lattice accounts for the measured
condensation energy ($\sim 3$ K/u.c.) \cite{PWA2} the best: $2 N^2 \times 3$ $K=96 $K
which is fairly close to $T_c=93$ K.
The generally quoted $\xi_{ab} \approx 15 \AA$ \cite{Tinkham,Welp} is indeed in accordance, with the $4 \times 4$ to $5 \times 5$ coherence
area.

\section{The calculated dielectric constant}

  In Table I. we have calculated the static dielectric function $\epsilon_c$ using Eqs. ~(\ref{epsc3}) and ~(\ref{epsc4}).

 It must be noted that experimental measurement of $\epsilon_c$ in cuprates is often troublesome. Different studies often
lead to different values due to the inaccuracy provided by reflectance or conductivity measurements.  
Measurements on polycrystalline samples of $Bi_2Sr_2RCu_2O_8 (R=Sm,Y)$ provided $\epsilon_c$
as large as $10^4$ at 300 K \cite{Takayanagi}.
Also, other studies indicate large $\epsilon_c$ in the parent insulators of HTSCs \cite{Takayanagi}.
It is even more difficult to accurately measure $\epsilon_c$ for a SC sample \cite{Takayanagi}.
In other studies also large $\epsilon_c$ values have been reported for $YBa_2Cu_3O_{6+\delta}$
and extremely large for the non-supercondcuting $PrBa_2Cu_3O_{6+\delta}$ \cite{Rey,Mazzara}.
Tsvetkov {\em et al.} reported the value of $\epsilon_c=11.3$ for Tl2201 using optical measurements \cite{Tsvetkov}.
$\epsilon_c$ can be extracted from the $c$-axis optical measurements using the
relation \cite{Tsvetkov,PWA2}
\be
\sqrt{\epsilon_c} = \frac{c}{\omega_p \lambda_c}, 
\ee
where $c, \omega_p$ and $\lambda_c$ are the speed of light, $c$-axis plasma frequency
and the $c$-axis penetration depth. 
In other cases only the $\epsilon_c$ of certain elements of the ion-core spacer is available
such as e.g. BaO in Hg1201 and in YBCO. 
It must be emphasized that the reported experimental $\epsilon_c$ values are frequency dependent
and show relaxation behaviour therefore the comparison is not easy with our reported values.
For the prototypical cuprate LSCO we get the value of $\epsilon_c=27.3$ which is
comparable with the experimental value of $23$ \cite{epsilonc} using $N=7$
which corresponds to $6 a_0 \approx 23.4 \AA$ ($\xi_{ab} \approx 20-30 \AA$ \cite{coher}).

\begin{table}
\caption[]
{Calculated dielectric constant $\epsilon_c$ using Eqs ~(\ref{epsc3}) and ~(\ref{epsc4})
in the single-layer cuprates.
}
{\scriptsize
\begin{tabular}{lccccccc}
 & $d (\AA)$ & $T_c (K)$ & $N$ & $\epsilon_c/Eq.~(\ref{epsc3})$ & $\epsilon_c/Eq.~(\ref{epsc4})$ & $\epsilon_c^{exp}$  \\ \hline
CaCuO2$^a$ &       4.64 &  89 &  5 & 147.9 & 100.0 &   \\ 
CaCuO2$^b$ &       4.64 &  110 & 5 & 96.7 &  80.8  &    \\ 
CaCuO2     &       3.19 &  89 &  8 & 123.5 &  83.5 & 
    \\ 
     &        &   &  9 & 94.9 &  64.1 & 
\\
 LSCO &      6.65 &  39 & 6 &  39.9 & 27.9 & $23 \pm 3^c$  \\
      &       &  & 7 & 39.2 & 27.3 &   \\
      &       &  & 8 & 16.3 &  11.3 &   \\
 Hg1201    &     9.5   & 95  & 5 & 17.7 & 27.6 & 34$^d$   \\
 Hg- (10 GPa) & $\sim 8.5$   & $\sim 105.$ & 5 & 15.9 & 27.6 & \\ 
 Hg- (20 GPa) & $\sim 8.2$   & $\sim 120.$ & 5 & 12.6 &  25.0 &   \\
 Tl2201 &    11.6 &  85 &  4 & 13.5  &  8.7 & 11.3$^e$ \\
        &         &     &  5 & 41.6  & 26.8  &  \\
  & $39.4^f$ \\
 Bi2201 & 12.2 & 20  & 4 & 214.5 & 32.6 & $\sim 40^g$ \\ 
  &  &   & 5 & 729.3 & 110.8 &  \\ 
  &  &   & 6 & 66.6 & 10.1 & \\  
  YBCO  &   $8.5^h$  & 93  & 4 & 27.5 & 19.4 & $34^f$ \\ 
        &            &      & 5 & 44.1 & 31.1 &  \\ 
 $LSCO^i$ &      6.65 &  51.5 & 7 & 21.6 & 20.2  & \\
\end{tabular}}
{\small
where $N$ is the number of lattice sites along the width of the square
superlattice ($N \approx \xi_{ab}/a+1$),
$dq$ is the charge separation in the charge ordered state, $d$ is the inter-layer distance in $\AA$ \cite{Mills}, $T_c$ is the experimental critical temperature.
$^a$ field effect doped material \cite{Schon}.
$^b$ $Ca_{0.3}Sr_{0.7}CuO2$ \cite{Azuma}, 
$^c$ \cite{epsilonc},
$^d$ experimental $\epsilon_c^{exp}$ values are taken from {\em Am Inst. of Phys. Handbook}, McGraw-Hill, Ed. D, E. Gray (1982),
the $\epsilon_c$ value of the ionic-background (Hg1201: BaO, Tl2201: $Tl_2O_5$),
The pressure dependent $T_c$ values and IL distance date are taken from \cite{Novikov,Gao},
$^e$ from \cite{Tsvetkov}, 
$^g$ from \cite{Kitano},
$^h$ for YBCO the reduced IL $d=8.5 \AA$ is used instead of the
$c$-axis lattice constant (inter-bilayer block distance),
$^f$ from \cite{Ariosa},
$^i$ from \cite{Bozovic}.
The notations are as follows for the compounds:
LSCO ($La_{0.85}Sr_{0.15}CuO_4$), Hg1201 ($HgBa_2CuO_{4+\delta}$), Tl2201 ($Tl_2Ba_2CuO_6$), 
 Bi2201 ($Bi_2Sr_2CuO_6$)
and YBCO ($YBa_2CuO_7$).
}
\end{table}


\begin{figure}[hbtp]
\begin{center}
\includegraphics*[height=4.5cm,width=6.5cm]{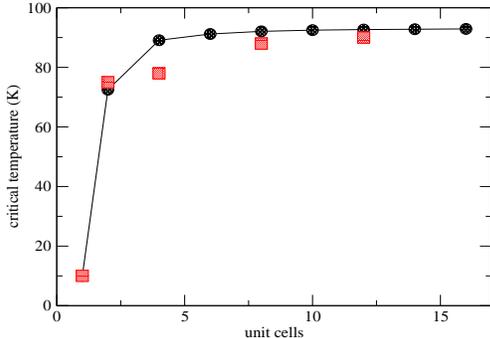}
\caption[]{The critical temperature $T_c$ (K, Eqs. ~(\ref{kbtc})-(\ref{epsc4})) {\em vs.} 
the number of unit cells along the $c$-axis in YBCO using
the $4 \times 4$ model.  Circles and squares correspond to
the calculated and experimental values \cite{Li}.
}
\label{fig_4}
\end{center}
\end{figure}


 We try to explain the doping function $T_c(\delta)$ using the $N \times N$.
 The coherence area ($\sim \xi_{ab}$) systematically varies with doping
 and reaches its minimum at optimal doping ($n_h \approx 0.16$).
 Therefore, the contraction of the pair-condensate wave-function $\Psi(r_1,r_2)$
 is the most pronounced at optimal doping accomodated by $4 \times 4$ to
 $5 \times 5$ superlattice.
When the system is {\em underdoped}, the $E_{el}^c[n_h(\delta)]$ scales almost linearily with $\delta$.
This is because the smaller $n_h(\delta)$ leads to smaller inter-layer charging energy
while $\epsilon_c$ is nearly unchanged.
This behaviour is clearly described by Eq.~(\ref{kbtc2}).
However, in the {\em overdoped} regime more and more charge is confined to the inter-layer
space, which increases $\epsilon_c$, hence decreases $T_c$.
The physical explanation is that the increased density of inter-layer hole-charge
screens the dielectric medium (e.g. $Ca^{2+}$), which leads to less effective polarization
of the condensed part of the hole charge.
In the {\em optimally doped} materials the confinement of the hole content in the planes is the
most pronounced in the SC state.
Doping induced out-of-plane charge transport, such as the strong c-axis anisotropy of the hole
content, 
leads to charge inhomogeneity, and might lead to large $\epsilon_c$ \cite{Takayanagi}.

 {\em Multilayer} and {\em pressure} effects on $T_c$ can also be discussed in terms of 
in-plane and out-of-plane localization of the hole charge.
In multilayer systems $E_{el}^c$ is composed of intra- and inter-block contributions.
We use the notation block for the multilayer parts $CaCu_2O_4$, $Ca_2Cu_3O_6$, etc. 
Eq. ~(\ref{gain_sc}) can then be modified for multilayer cuprates as follows,
\be
 E_{cond} \approx E_{el}^{c,intra} + E_{el}^{c,inter},
\label{multil}
\ee
where
\be
E_{el}^{c,intra}=(l-1) E_{el}^c(d_{intra}, \epsilon_{intra}),
\ee
l denotes the number of layers.
Therefore, intra-block charging energy further enhances $T_c$ on top of the corresponding
single-layer inter-block value.
YBCO is a peculiar example of cuprates in which HTSC is purely coming from {\em inter-block
coupling} as it was explained on the basis of $YBa_2Cu_3O_{7-\delta}/PrBa_2Cu_3O_{7-\delta}$
superlattice structures \cite{Plakida}.
In YBCO the $Y(CuO_2)_2^{-2}$ bilayer alone does not show HTSC when isolated from each other
in artificially layered materials ($T_c \approx 20 K$).
The dependence of $T_c$ in YBCO on the number of unit cells along the $c$-axis
is calculated using Eq. (25) and the results are depicted on FIG 4.
The experimental points \cite{Li} are relatively well reproduced
indicating that $\sim 5$ unit cell thick thin film is already the reminiscent of
the bulk properties of cuprates.

 It is worth mentioning the {\em intercalation} experiment on Bi2212 \cite{Choy}.
Intercalation of I or organic molecules into the $(BiO)_2(SrO)_2$ layers of Bi2212 results
in no significant change in $T_c$.
Basically inter-layer intercalation is introduced to reduce interlayer coupling (test of inter-
layer theory, ILT) and to enhance anisotropy in SC properties.
It was concluded that these results presents great challenge to ILT \cite{PWA} and support
low-dimensional SC theories.
In the light of the $N \times N$ model, it is not surprising that in Bi2212 the nearly isolated bilayers are 
superconductors.
The single layer material Bi2201 ($Bi_2Sr_2CuO_6$) gives very low $T_c$ ($\approx 20 
K$) \cite{Konstantin},
which indicates that the dielectrics $(BiO)_2(SrO)_2$ strongly reduces IL coupling, indeed
the estimated
$\epsilon_c$ is quite large (Table I) which is attributed to the weakly interacting $(BiO)_2$
bilayer (the BiO-BiO distance is $\sim 3.7 \AA$ \cite{Choy}). 
Dielectric constant measurements \cite{Varma} on Bi-compounds also provide relatively large
dielectric constants in accordance with our calculations.
The extremely large conduction anisotropy $\gamma \sim 10^6$ found in Bi-based cuprates also supports these findings \cite{Plakida}.
Nearest-neighbouring Cu-O planes in Bi-compounds are nearly insulated \cite{Ariosa}.
Therefore, in the multilayer Bi-compounds bilayer and trilayer blocks are responsible for
the high-$T_c$, inter-block coupling is negligible.
The multilayer Bi-compounds provide then an example for {\em pure
intra-block} HTSC. In these materials the multilayer-blocks are dielectrically isolated.
However, in the most of the cuprates $T_c$ is enhanced both via intra- and inter-block effects.
An important consequence of the bilayer $N \times N$ model is that {\em isolated single layers do not show HTSC},
coupling to next-nearest $CuO_2$ plane is essential in HTSC.
It is possible then to estimate $\epsilon_c$ for the bilayer block, since $E_{el}^{c,inter} \approx 0$ in the Bi-compunds in  
Eq.~(\ref{multil}).
In the hypothetical bilayer system in Table 1. ($Ca(CuO_2)_2, d=3.19 \AA, T_c=89 K$),
which is the building block of multilayer cuprates,
we find the value of $N=8$ accounting for realistic
$\epsilon_c$. $N=8$ is in accordance with the measured relatively
"large" coherence length of $\xi_{ab} \approx 27 \AA$ ($(N-1)a_0=7 a_0 \approx 27.3 \AA$) \cite{Bi_coher}.

 Also, upon pressure (p) the inter-layer spacing decreases, which increases $E_{el}^c$ without
the increase of the hole content. Saturation of $T_c(p)$ is reached simply when 
increasing IL sterical repulsion starts to destabilize the system.  
In systems, such as YBCO or LSCO, negative or no pressure dependece of $T_c$ is found
\cite{Plakida} due to the short Cu-O apical distance ($d_{CuO} \approx 2.4 \AA$) which leads to steric repulsion at ambient pressure and
to the weakening of HTSC. In these systems the net gain in charging energy is not enough 
to overcome steric repulsions.

 With these results, we are now in a position to reach the conclusion that the $N \times N$ model can readily account for at least certain physical properties
of multilayer cuprates, such as pressure,  doping and multilayer dependence of $T_c$.
Furthermore it is possible to make some estimations on the upper limit of $T_c$
using Eq. ~(\ref{epsc4}) for $T_c$. Assuming relatively small $\epsilon_c \approx 10$ and short inter-layer
spacing $d \approx 7.0 \AA$ we get the value of $T_c \approx 333 K$ for a 
strongly localized electron pair with $5 \times 5$ coherence area. Of course, we have no clear cut knowledge at this moment
on how the pair condensate wave-function spreads upon varying $\epsilon_c$ and $d$.
Effective localization of the Cooper-pair wave-pocket could lead, however, to room temperature
superconductivity under proper structural and dielectric conditions if the model presented
above is applicable.

 According to the stripe scenario, the charge ordered state of the coherence area can also be seen 
 in many cuprates as one-dimensional stripe order or charge density waves
 \cite{Howald}.
 The striped antiferromagnetic order found in LSCO by magnetic neutron 
 scattering experiments \cite{Lake} also implies a spin-density of periodicity $\sim 8 a$
 a reminiscent of the coherence area. 
 The incommensurate "checkerboard" patterns seen with a spatial periodicity of $\sim 8a_0$ in the vortex core of Bi2212 obtained by scanning tunneling microscopy \cite{Hoffman} is also consistent
 with our $N \times N$ hole-anti-hole charge ordered state where $N=8$ to $9$ in BSSCO. 

 Finally we mention the recent results of Bozovic {\em et al.} \cite{Bozovic}
obtained for LSCO thin films under epitaxial strain. They reached the
record $T_c=51.5 K$ for 15-unit-cell thick film of LSCO on $LaSrAlO_4$ substrate.
The small variation of the $ab$- or $c$-axis lattice constants in our model accounts for only
$1-2$ K increase in $T_c$.
We explain the more then $10$ K enhancement of $T_c$ with the decrease of $\epsilon_c$
(Table I.) under the conditions they used ($O_3$ annealing, epitaxial strain provided by the substrate).



\section{Conclusion}

 In this paper we studied the pair condensation and confinement of the hole-content on a $CuO_2$ superlattice layer
as a function of inter-layer distance and dielectric permittivity of the charge reservoir.  

\begin{itemize}
\item The assumption of a $2D \Leftrightarrow 3D$ quantum phase transition of the hole-content
at $T_c$
in HTSC materials is thought to be an important general feature of pair-condensation and is supported
by $c$-axis optical measurements and by first-principles calculations.
Our proposal is that the $c$-axis charge dynamics of the hole-content contributes
significantly to the condensation energy below $T_c$.
We find that
the inter-layer capacitance is temperature dependent in cuprates and therefore the
drop of the $c$-axis dielectric constant can be seen below $T_c$.
This is what leads to then the stabilization of the superconducting state {\em vs.}
the normal state.
\end{itemize}

\begin{itemize}
\item We have found that a pair condensate can be distributed on a $N \times N$ square lattice layer
in such a way that the lattice sites are filled by $\pm q=[0,0.16e]$ condensed charge alternatively depending on the hole($+q$)-anti-hole($-q$) charge separation $dq$.
In this way a charge ordered state of the pair-condensate occurs 
with a "checkerboard" like pattern seen recently by experiment \cite{Hoffman}.
The phase separation of hole-electron pairs
(hole-anti-hole pairs) in this model is stabilized electrostatically.
The maximum charge separation is $dq=0.32e$, if the optimal hole content $n_h \approx 0.16e$.
\end{itemize}

\begin{itemize}
\item In the adjacent layer the electron-hole pairs are distributed in a complementer way (charge asymmetry) in order to maximize the inter-layer charging energy. 
Holes on one plane are covered by anti-holes on the other, and vice versa.
In this way we derived a charge ordered $N \times N$ bilayer model of the superconducting state
with inter-layer charge mirror symmetry which directly leads to inter-layer Coulomb energy 
gain in the superconducting state.
\end{itemize}

\begin{itemize}
\item The superlattice nature of the pair condensate is directly related to the
smallest size of the condensate wave-pocket which is remarkably comparable with
the measured in-plane coherence length of $\xi_{ab}=10-20 \AA$  ($4a_0 \approx 15.6 \AA$)
of cuprates, where $a_0 \approx 3.9 \AA$ is the in-plane lattice constant.
The coherence area of the Cooper pair in cuprates is strongly localized,
which is due to inter-layer charging effects.
\end{itemize}

\begin{itemize}
\item The bilayer model with $4e$ boson charge naturally implies the mass enhancement of $m^{*} \approx 4 m_e$
in accordance with measurements.
The experimental condensation energy obtained for Tl2201 and YBCO is also in accordance with
the $4 \times 4$ to $5 \times 5$ bilayer superlattice model.
\end{itemize}

\begin{itemize}
\item
The static dielectric constant $\epsilon_c$ is calculated for a couple of cuprates
and compared with the available
experimental measurements. The general agreement is quite good indicating that
the pure inter-layer electrostatic model leads to proper description of
the static dielectric response of these layered materials. 
\end{itemize}

\begin{itemize}
\item The basic microscopic mechanism of HTSC is to be understood within the 
BCS-Eliashberg theory at low temperatures. The limiting critical temperature
for BCS-type superconductor is around $20 K$ as it was found for cuprates
with nearly isolated $CuO_2$ layers (BSCCO, YBCO/PrBCO superlattices etc.).
At higher temperatures
up to $T_c$ the SC phase is stabilized {\em vs.} the NS by the inter-layer Coulomb
interaction due to the inter-layer charge antisymmetry of the pair condensate.
\end{itemize}

\begin{itemize}
\item We propose to understand chemical- and field-effect doping as a hole-content confinement mehanism,
induced by the electrostatic field of the charged dopand atom (chemical doping) or by the external field (in a field effect transistor)
below $T_c$.
\end{itemize}

The detailed study of this model showed that the inter-layer charging energy is proportional
to the thermal motion at $T_c$, if the $c$-axis dielectric constant $\epsilon_c$ and the coherence area is appropriately
chosen.

  If the physical picture derived from our model is correct, it should be a guide for 
further experimental studies aiming to improve SC in cuprates or in other materials.
This can be done by tuning the IL distance and $\epsilon_c$ (increasing the polarizability of the dielectric, hence decreasing $\epsilon_c$) in these materials.
The application of this model to other class of HTSC materials, such as fullerides or
$MgB_2$ is expected to be also effective.

 Further development of the model presented here, such as using charge density $\rho(r)$ instead of
point charges within a self-consistent quantum theory, would make it possible to study the external field or chemical doping induced $c$-axis charge
distribution in cuprates. This is however, beyond the scope of the present article
and is planned in the near future.

\section{acknowledgement}

{\small
It is a privilige to thank M. Menyh\'ard for his continous support.
I greatly indebted to E. Sherman for reading the manuscript carefully
and for the helpful informations.
I would also like to thank for the helpful discussions with T. G. Kov\'acs.
This work is supported by the OTKA grant MFA-42/2002
from the Hungarian Academy of Sciences}
\\


\end{document}